\documentclass[review]{elsarticle}

\usepackage{lineno,hyperref}
\usepackage{amssymb, amsmath}
\usepackage{lipsum}
\usepackage{caption}

\modulolinenumbers[5]

\journal{Big Data Research}

\begin{document}

\newcommand{\antiInstanton}{(anti\nobreakdash-)instanton}
\newcommand{\AntiInstanton}{(Anti\nobreakdash-)Instanton}
\newcommand{\latticeQCD}{lattice QCD}
\newcommand{\LatticeQCD}{Lattice QCD}
\newcommand{\jcn}{JCN}
\newcommand{\quotes}[1]{``#1''}
\newcommand{\tcd}{topological charge density}
\newcommand{\Tcd}{Topological charge density}
\newcommand{\imagi}{\mathsf{i}}
\newcommand{\su}{\mathsf{SU}}

\graphicspath{
	{./graphics/}, 
}

\begin{frontmatter}

\title{Joint Contour Net Analysis for Feature Detection in Lattice Quantum Chromodynamics Data}


\author[swansea_cs,swansea_phys]{Dean P.Thomas}
\ead{798295@swansea.ac.uk}

\author[kcl]{Rita Borgo}
\ead{rita.borgo@kcl.ac.uk}

\author[swansea_cs]{Robert S.Laramee}
\ead{r.s.laramee@swansea.ac.uk}

\author[swansea_phys]{Simon J.Hands}
\ead{s.j.hands@swansea.ac.uk}

\address[swansea_cs]{Department of Computer Science, Swansea University, Swansea, United Kingdom}
\address[swansea_phys]{Department of Physics, Swansea University, Swansea, United Kingdom}
\address[kcl]{Department of Informatics, Kings College London, London, United Kingdom}

\begin{abstract}
	In this paper we demonstrate the use of multivariate topological algorithms to analyse and interpret Lattice Quantum Chromodynamics (QCD) data.	Lattice QCD is a long established field of theoretical physics research in the pursuit of understanding the strong nuclear force.  Complex computer simulations model interactions between quarks and gluons to test theories regarding the behaviour of matter in a range of extreme environments.  Data sets are typically generated using Monte Carlo methods, providing an ensemble of configurations, from which observable averages must be computed.  This presents issues with regard to visualisation and analysis of the data as a typical ensemble study can generate hundreds or thousands of unique configurations.

	We show how multivariate topological methods, such as the Joint Contour Net, can assist physicists in the detection and tracking of important features within their data in a temporal setting.  This enables them to focus upon the structure and distribution of the core observables by identifying them within the surrounding data.  These techniques also demonstrate how quantitative approaches can help understand the lifetime of objects in a dynamic system.
\end{abstract}


\end{frontmatter}


\section{Introduction}

\noindent
Recent advances in multivariate topological visualisation have provided new approaches for detecting interesting phenomena in scalar fields of more than one variable.  Much of this work builds upon existing techniques used to understand the topology of scalar fields \textemdash{} where critical events, such as the creation and merging of unique features, are captured using graph structures.  The ability to examine multiple fields in parallel also presents a method for tracking objects in higher dimensional data sets.

In this paper we use the joint contour net (\jcn{}) algorithm~\cite{Carr2013} to track objects with a finite lifetime across multiple time steps.  \AntiInstanton{}s are 4D \quotes{pseudo-particles} studied by domain scientists that are localised to specific locations in 4D Euclidean space-time.  Existing statistical physics methods are able to predict the existence of these objects; however, more complex properties such as their structure and lifetime are more difficult to evaluate.  

To demonstrate how multivariate topological visualisation techniques can benefit \latticeQCD{} scientists we focus upon analysing a single instanton pre-identified by existing physics methods.  Due to the Euclidean nature of \latticeQCD{}, where space and time are treated equivalently, the techniques used in this paper can also be used to scan volumes with a temporal component ($xyt$, $xzt$, $yzt$) along a spatial axis.  Whilst not seen as a direct replacement for viewing these fields in their native 4D embeddings, this technique presents an interesting approach to pin-point critical events in the topology of a single evolving field.  

By carrying out this case study we intend to answer the following questions:

\begin{itemize}
	\item Can we use the \jcn{} to track an instanton between multiple neighbouring time slices? 
	\item Can extra properties about instantons be determined through multivariate persistence measures?
	\item Can the Reeb skeleton be used to simplify the field and summarise properties of observables?
\end{itemize}

The remainder of this paper begins with an overview of the necessary background information in Section~\ref{sec::background}.  Section~\ref{sec::temporal_jcn_implementation} briefly introduces the application used to carry out the work in this paper.  We then give a description of the observed data in Section~\ref{sec::multivariate_temporal_visual_analysis}, this forms the basis of a quantitative approach to evaluating the topological structure of the data in Section~\ref{sec::multivariate_temporal_quantitative}.  We then show an experimental use of the \jcn{} to locate and study the structure of an \antiInstanton{} within an entire 4D data set in Section~\ref{sec::multivariate_temporal_full_lattice}.  The paper is concluded in Section~\ref{sec::multivariate_summary_temporal} where we summarise our findings from the case study.
\section{Background}
\label{sec::background}

\noindent
In this section we introduce the relevant background information required to read the rest of the paper.  We begin by providing an overview of multivariate topological visualisation algorithms, before introducing aspects of \latticeQCD{} that are relevant to this paper.  Much of multivariate topology uses a generalisation of univariate topology concepts where structures such as the Reeb graph~\cite{shinagawa1991surface} summarise the topology of a scalar field.  For a more in depth overview of the use of topology in visualisation we refer the reader to the survey paper by Heine et al.~\cite{heine2016survey}.

\subsection{Multivariate topological visualisation}
\label{sec::multivariate_topological_visualisation}

\noindent
The \emph{Reeb space} is a generalisation of the Reeb graph enabling multivariate or temporal data to analysed.  The first discussion of using the Reeb space to compute topological structure of multiple functions is presented by Edelsbrunner et al.~\cite{edelsbrunner2008reeb}, where it is suggested that the Reeb space can be modelled mathematically in the form $f : \mathbb{M} \mapsto \mathbb{R}^{k}$, where $\mathbb{M}$ represents the domain and $f$ the output of $k$ scalar functions.  For the simple case, where $k = 1$, this is directly comparable to the Reeb graph.  The Reeb space extends this formulation to situations where $k \ge 2$.

\subsubsection{The joint contour net}
\label{sec::the_joint_contour_net}

\begin{figure*}
	\centering
	\includegraphics[width=.96\textwidth]{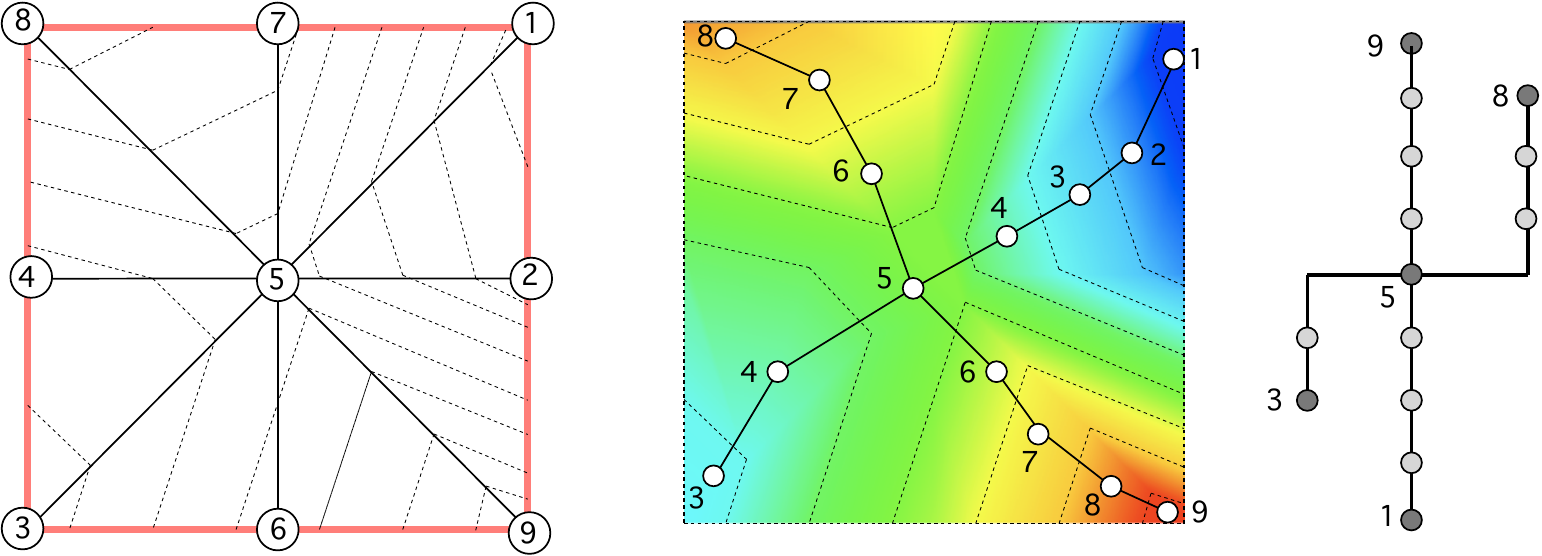}
	\includegraphics[width=.96\textwidth]{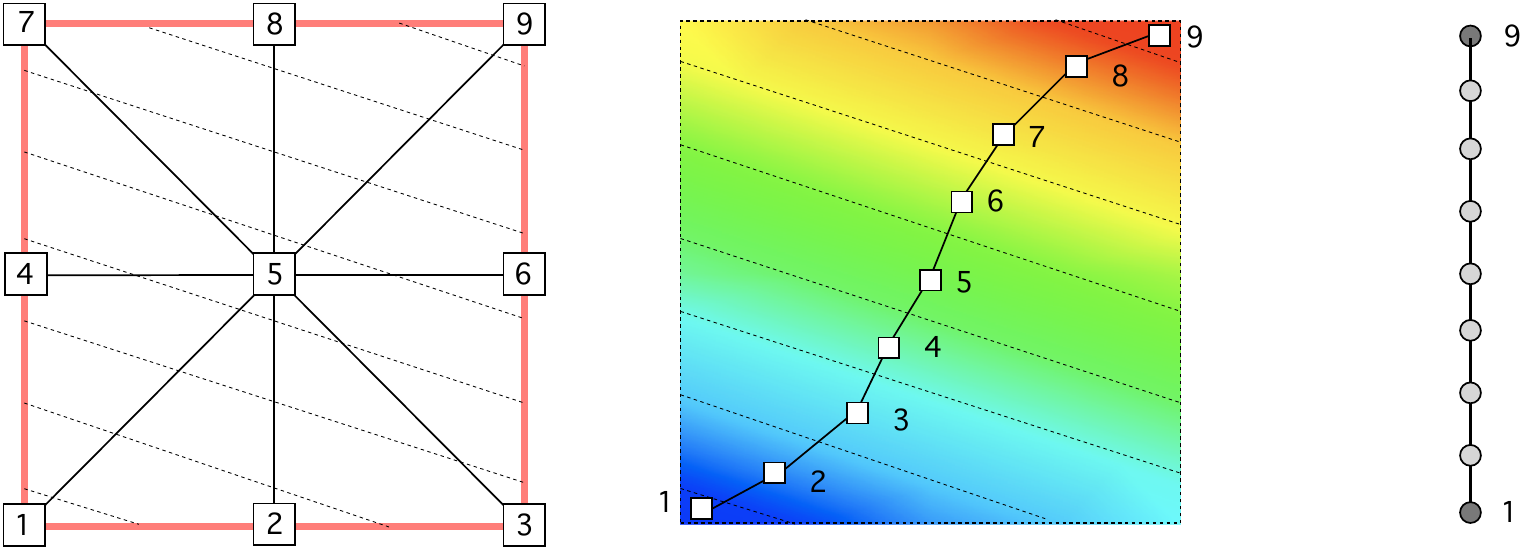}
	\footnotesize{Image taken from Duke et al.~\cite{duke2012visualizing}}
	\caption{Two simple scalar functions defined on a simplicial grid (left) where the dotted lines represent the quantisation intervals.  The quantised contour tree for each function (right) is shown mapped to scalar field in the centre.}
	\label{fig::jcn_input}
	\includegraphics[width=.5\textwidth]{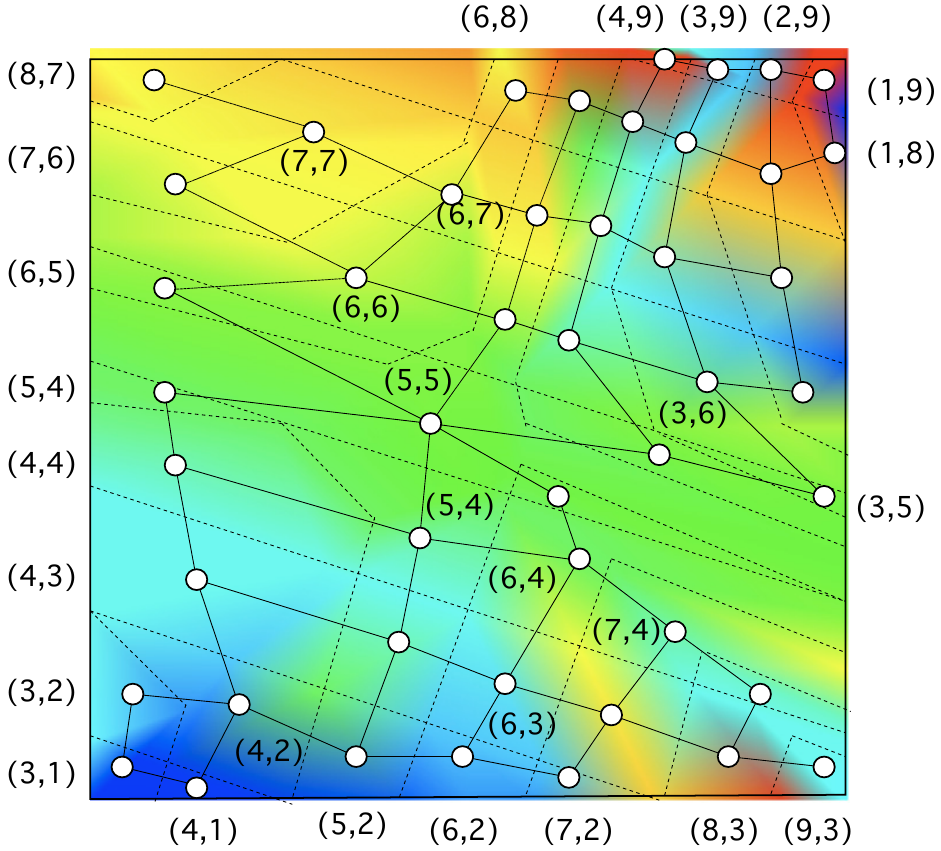}\\
	\footnotesize{Image taken from Duke et al.~\cite{duke2012visualizing}}
	\caption{The \jcn{} capturing the bivariate topology of the two simple functions shown in Fig.~\ref{fig::jcn_input}.  The bivariate field is decomposed by overlaying the quantisation intervals of the two input fields (dotted line).  A vertex is placed at the barycentre of each region, or joint contour slab, and edges mark adjacency.}
	\label{fig::jcn_output}
\end{figure*}

\noindent
Carr et al.~\cite{Carr2013} presented the first discrete representation of the Reeb space using the \emph{Joint Contour Net} (\jcn{}).  For functions of $n$ variables defined in an $\mathbb{R}^m$ dimensional space the algorithm approximates the Reeb space as a number of multivariate contours named joint contour slabs.  These represent connected regions of the domain with respect to the isovalue of multiple functions.  In situations where $n \geq m$ the \jcn{} can still be computed; however, the output is not an approximation of the Reeb space but instead a subdivision of the input geometry over $n$ variables.  The \jcn{} captures the Reeb space as an undirected graph structure, where vertices represents slabs of $n$ isovalue tuples, and edges are used to show adjacency between regions.  An example \jcn{} of two scalar functions is presented in Figures~\ref{fig::jcn_input} and~\ref{fig::jcn_output}.  The \jcn{} has previously been used to study multivariate data originating from nuclear scission simulations~\cite{duke2012visualizing, schunck2014description, schunck2015description} and hurricane measurements~\cite{geng2014visual}.

The Reeb skeleton is a simplified graph structure that takes into account the size of connected components, allowing measures of persistence to be assigned to its arcs to aid multivariate simplification~\cite{chattopadhyay2015multivariate}.  Lip pruning techniques, similar to the leaf pruning method of simplification found in univariate topological structures~\cite{carr2004simplifying} can then be applied to progressively remove noisy features in the multi-field.  Example persistence measures applied to the \jcn{} include the accumulated volume of joint contour slabs in a connected region.  Alternatively, the Reeb skeleton can be used to quantify regions of the multivariate topology for analysis.


\subsection{Lattice Quantum Chromodynamics}
\label{sec::lattice_quantum_chromodynamics}

\noindent
Kenneth Wilson was the first physicist to suggest that a discrete 4D lattice could be used to model properties of quark-gluon fields~\cite{WIL74}.  The lattice is a hyper-torus in Euclidean space-time, meaning that the three spatial dimensions and the time dimension are treated as equal.  Periodic boundary conditions, where the minima and maxima on each axis are connected, are used so that it is impossible to consider any position on the lattice to be on a boundary.  The scale at which \latticeQCD{} acts is typically in the region of $2$ to $4$ femtometres ($10^{-15}$m), with state of the art simulations having a lattice spacing $a$ of $0.02$ to $0.04$ femtometres.

Quarks are located on the lattice at positions with integer indices referred to as \emph{sites}.  From each lattice site four \emph{link variables} are used to model the gluon potential in the $x, y, z$ and $t$ directions between two sites.  Each link variable is a member of the \emph{special unitary group} of matrices, identified using the notation $\su(n)$.  The value of $n$ represents the number of charge colours used in the gauge theory, with true QCD defined with $n=3$.  However, in this work we use a simplified two colour model of the theory using $\su(2)$ matrices.  Colour is used in this context to parametrise the concept of colour neutrality in a form similar to that of positive and negative charge.  One of the primary reasons for using a simplified model is that it allows us the freedom to vary the chemical potential of the system~\cite{Hands:2000ei}.  Chemical potential represents the energy change as either a quark is added or an anti-quark is removed from a system.  Varying the parameter enables exotic forms of matter, such as neutron star cores, to be simulated.

The discrete nature of the lattice means it is possible to calculate paths around sites in space and time.  Certain configurations of closed loops on the lattice are used to generate the scalar field observables of \latticeQCD{}.  The most basic unit closed loop in any directions is commonly referred to as a \emph{plaquette}; computing the average plaquette in all 4 dimensions produces the \emph{Wilson action} observable.  In this work we mainly focus upon the \tcd{} field, computed as a loop in all four space-time dimensions from each site~\cite{di1981preliminary}.  Regions of the lattice where the \tcd{} reach a minima or maxima indicate the presence of \antiInstanton{} pseudo-particles \textemdash{} one of the primary observables of \latticeQCD{}.  These have a finite extent in the time dimension and are able to appear and disappear, unlike real particles.  

In order to reveal the structure of \antiInstanton{} observables, the effect of quantum fluctuations must be minimised through a noise reduction technique known as \emph{cooling}.  After applying cooling what should remain are long range physical interactions that characterise the existence of an \antiInstanton{}; however, overly aggressive use of cooling can result in the destruction of the core observables.

\section{Implementation}
\label{sec::temporal_jcn_implementation}

\noindent
In this work we use the \jcn{} implementation supplied as part of the Multifield Extension of Topological Analysis (META) project~\cite{meta2015}.  This provides a number of filters for multivariate data that can be applied to the visualisation pipeline as part of a VTK~\cite{schroeder2006visualization} workflow.  Filters are included for creating an initial \jcn{} decomposition of the input fields that are able to be presented in graph form.  As part of the process of creating the \jcn{} individual joint contour slabs, in the form of polygon meshes, are created as the union of multiple smaller fragments.

Input fields are placed on to a common set of sampling points in three dimensions and each cubic cell is subdivided into 6 tetrahedra using a Freudenthal subdivison.  In order to handle periodicity an additional cell is constructed to link the minimum and maximum samples on each axis.

\subsection{Interactive user interface}
\label{sec::temporal_gui}

\begin{figure*}[!htb]
	\centering
	\includegraphics[width=.80\textwidth]{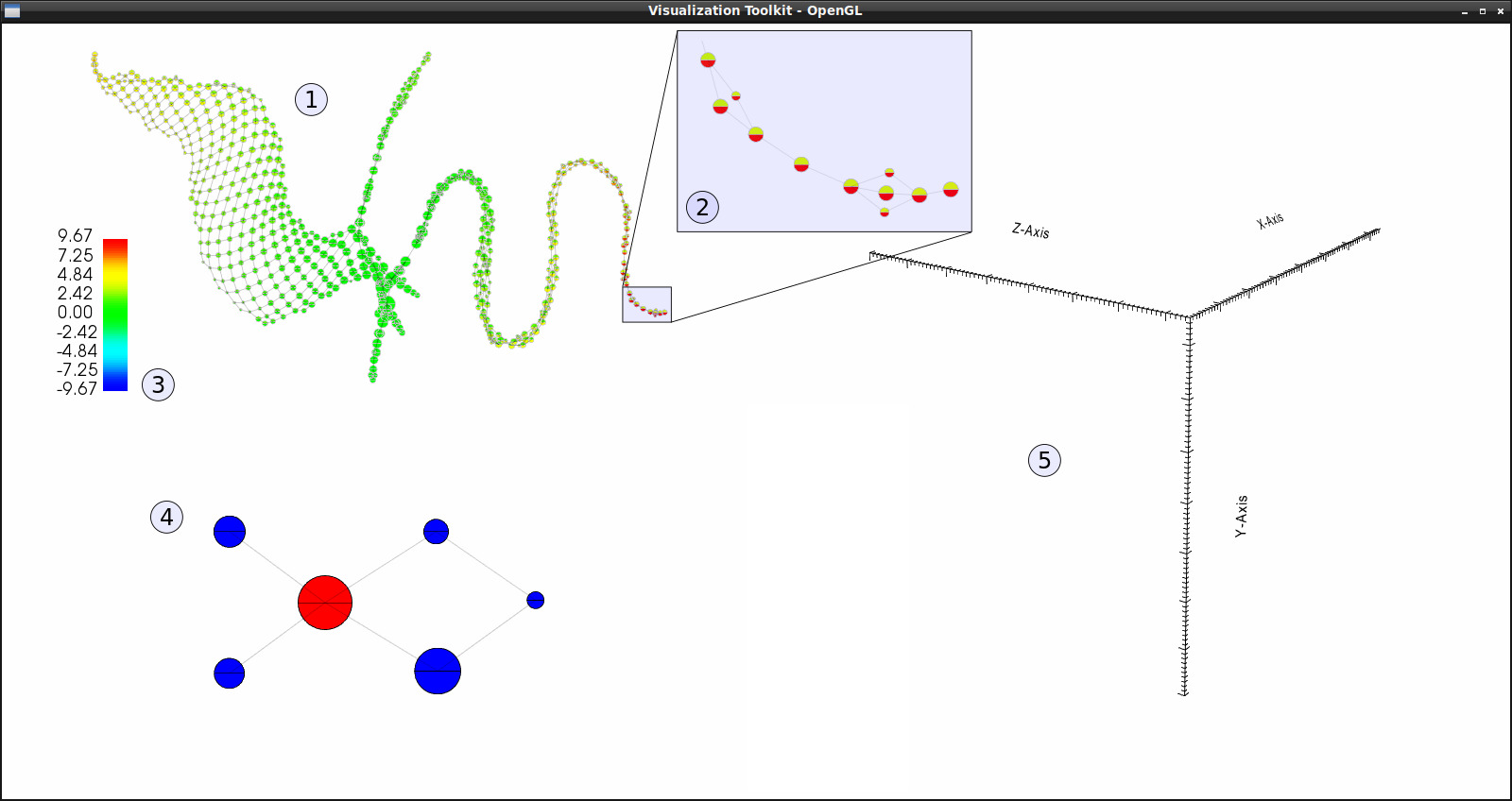}
	\caption{The user interface used to explore temporal \latticeQCD{} data as captured by the \jcn{}.  The \jcn{} captures the topology as a graph structure (1) \textemdash{} displayed here in a spring layout.  Each \jcn{} glyph (2) represents a slab (or quantised contour) using scale to provide feedback on the overall size of the slab.  Glyphs are coloured by isovalue (3), where the top half of each glyph represents the first input field and the bottom the second input.  Also visible is the Reeb skeleton (4) which captures the \jcn{} in a simplified form; red glyphs represent major changes in topological connectivity and blue glyphs represent relatively stable regions of topological structure.  Slabs are only rendered (5) to reflect any selections made by the user; in this view nothing has been selected.}
	\label{fig::vtk_tool_unselected}
\end{figure*}

\begin{figure*}[!htb]
	\centering
	\includegraphics[width=.80\textwidth]{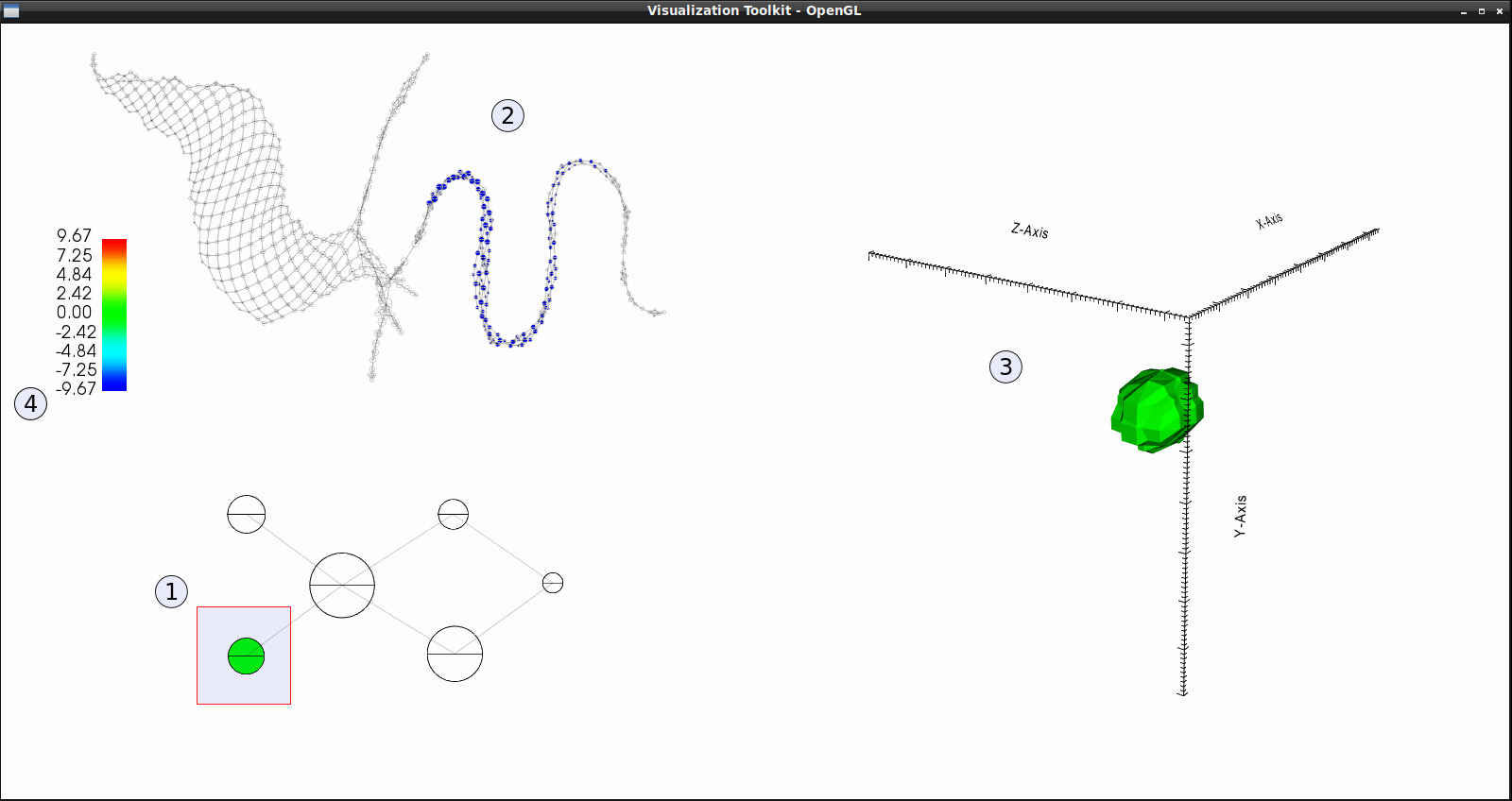}
	\caption{The user interface allows the user to select nodes in the Reeb skeleton (1) or \jcn{} (2) using rubber band selection.  Green glyphs in the Reeb skeleton correspond directly to the blue glyphs in the \jcn{} view.  Selected vertices are displayed as quantised contours (3) using a colour transfer function (4) determined by the bivariate input fields.}
	\label{fig::vtk_tool_selected}
\end{figure*}

\noindent
This case study was performed using a modified version of the interactive tool used by Geng et al. for the analysis of hurricane data~\cite{geng2014visual}.  Modifications of the software were largely made to facilitate collection of quantitative measurements for analysis.  In addition, the transfer functions used to colour the glyphs have been modified to enhance feedback for \latticeQCD{} fields that are often centred on zero~\cite{levkowitz1997, ware2012,ward2015}.  Figures~\ref{fig::vtk_tool_unselected} and~\ref{fig::vtk_tool_selected} give a visual overview of the user interface.

\section{Visual analysis of lattice objects in four dimensions.}
\label{sec::multivariate_temporal_visual_analysis}

\noindent
In the following section we give details of how an \antiInstanton{} can be located and tracked using the \jcn{}.  The entire workflow is given, beginning with the steps a domain scientist would use to locate a potentially interesting observable in a typical study.  We also discuss important considerations such as the slab size parameter which defines the resolution the scalar field is captured at.

\subsection{Selecting a candidate configuration for visual analysis}
\label{sec::multivariate_temporal_configuration_selection}

\noindent
Data used in this case study originates from a lattice with 16 spatial sites and 8 time sites, otherwise denoted as a $16^3 \times 8$ or \quotes{hot} lattice on account of the short temporal extent.  This particular lattice is chosen as it features relatively few time slices in comparison to other ensembles where the temporal extent can exceed $32$ steps.  The \jcn{} is demonstrated capturing a single topological object across the temporal axis including the periodic boundary between $t = 8$ and $t = 1$.  We also use the short temporal axis to push the limits of the \jcn{} by queuing multiple time slices on the multi-field to see if we are able to observe a signature for the entire 4D hyper-volume (Sec.~\ref{sec::multivariate_temporal_full_lattice}).

\begin{figure*}[!h]
	\centering
	\begin{minipage}{\textwidth}
		\includegraphics[width=\textwidth]{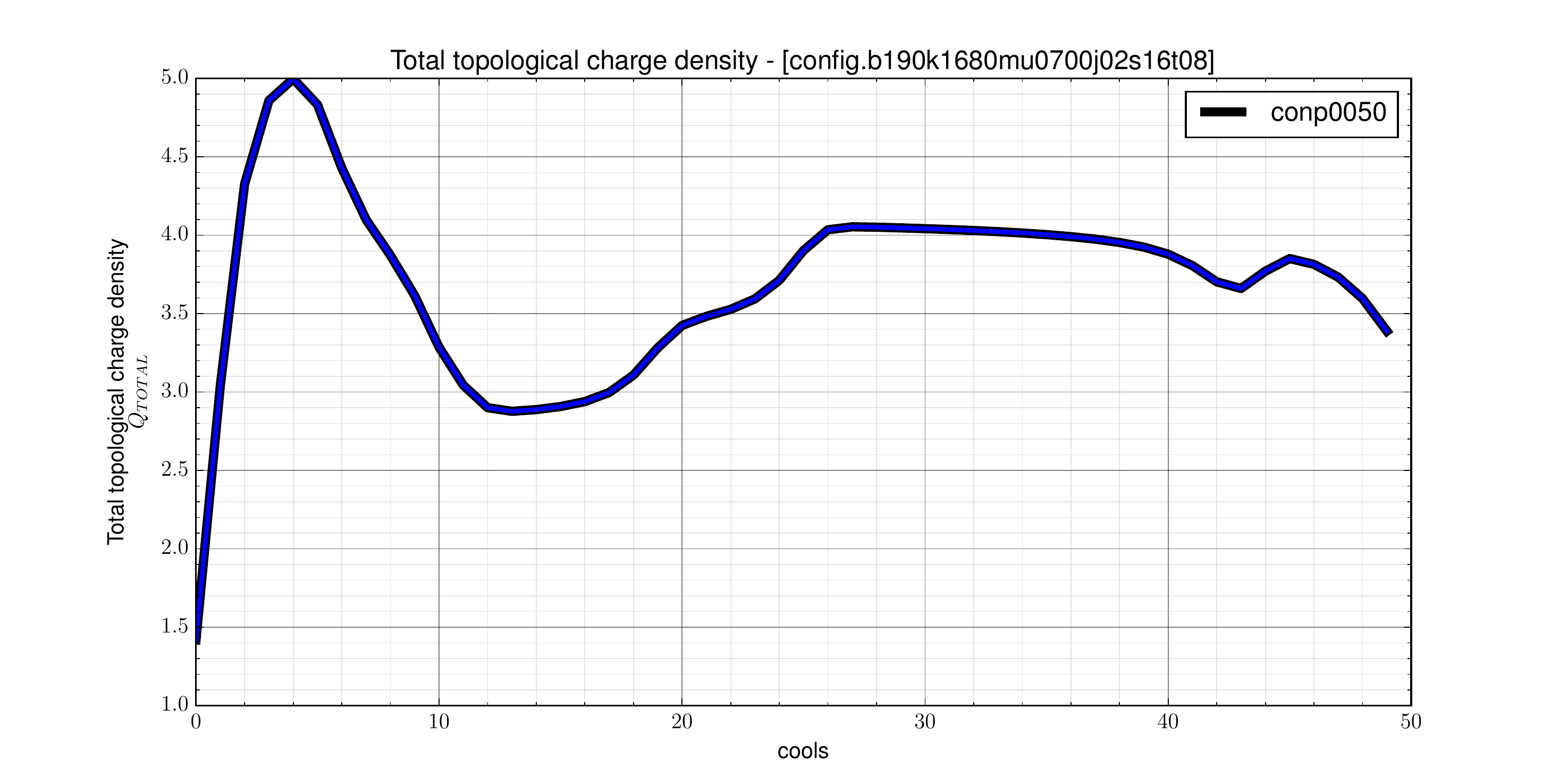}
		\caption{Total topological charge $Q_{TOT}$}
		\label{fig::topological_charge_total}
	\end{minipage}
	\hfill
	\begin{minipage}{\textwidth}
		\includegraphics[width=\textwidth]{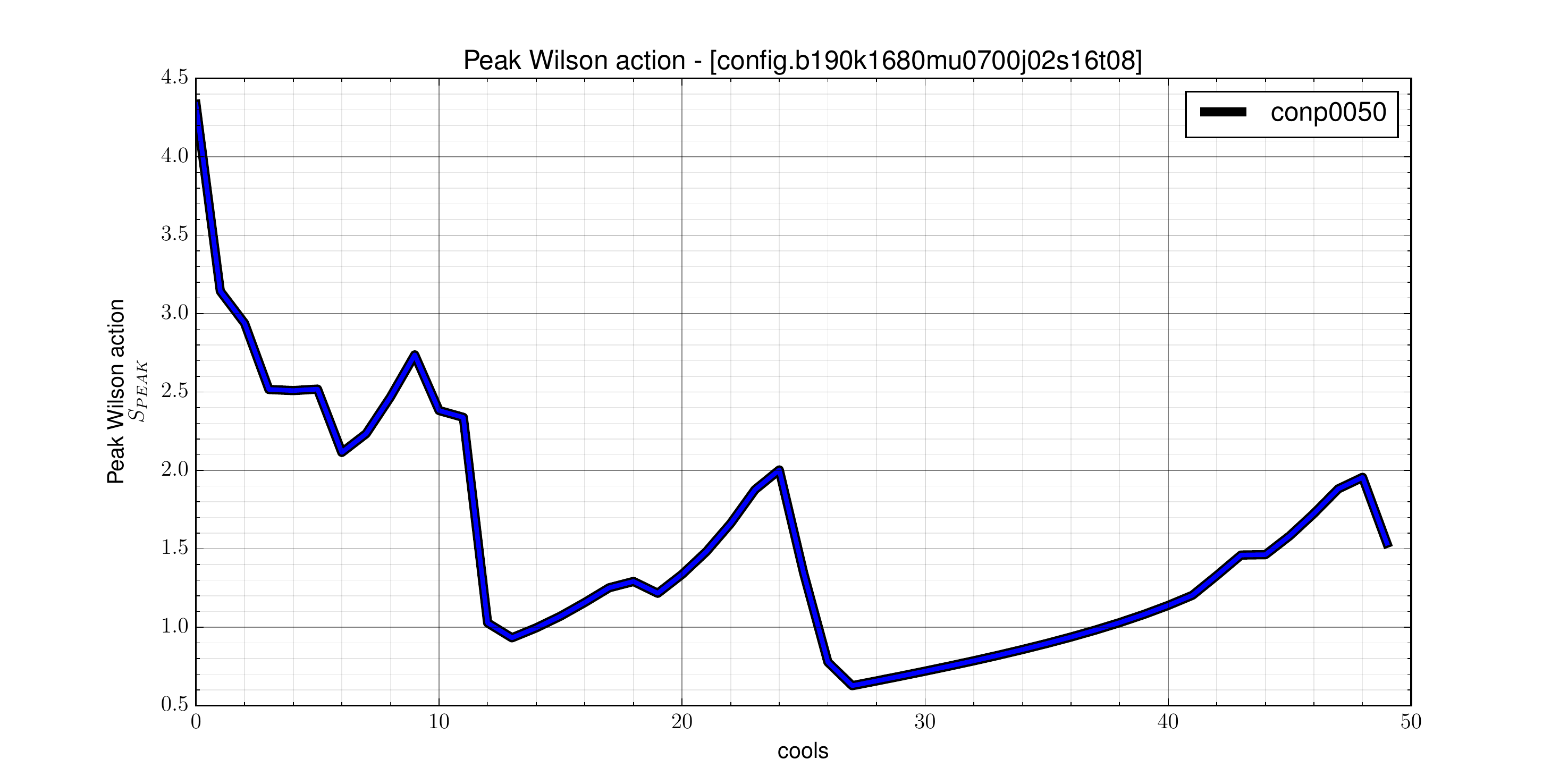}
		\caption{Peak Wilson action $S_{PEAK}$}
		\label{fig::wilson_action_peak}
	\end{minipage}
\end{figure*}

The configuration used in this study \emph{conp0050} originates from an ensemble with a chemical potential $\mu = 0.7$.  The lattice is pre-cooled for $30$ iterations to a stable state, validated by inspection of the total topological charge density and peak Wilson action graphs.  At this point in the cooling process the total topological charge $Q_{TOT}$ remains flat for many cooling iterations (Fig.~\ref{fig::topological_charge_total}) and the peak Wilson action $S_{PEAK}$ follows a smooth trajectory (Fig.~\ref{fig::wilson_action_peak}).

\begin{table*}[]
	\centering
	\begin{minipage}{0.48\textwidth}
		\resizebox{\textwidth}{!}{
			\begin{tabular}{lllll}
				Cools &	 $S_{MAX}$     & $Q_{MAX}$      & $Q_{MIN}$      &  \\
				\hline
				27    & (11, 4, 13, 5)  & (11, 4, 13, 5) & (6, 2, 7, 2) &  \\
				28    & \textbf{(9, 8, 10, 5)}  & \textbf{(9, 8, 10, 5)} & (6, 2, 7, 2) &  \\
				29    & (9, 8, 10, 5)  & (9, 8, 10, 5) & (6, 2, 7, 2) &  \\
				30    & (9, 8, 10, 5)  & (9, 8, 10, 5) & (6, 2, 7, 2) &  \\
				31    & (9, 8, 10, 5)  & (9, 8, 10, 5) & (6, 2, 7, 2) &  
			\end{tabular}}
		\end{minipage}
		\hfill
		\begin{minipage}{0.48\textwidth}
			\resizebox{\textwidth}{!}{
				\begin{tabular}{lllll}
					Cools	   &	 $S_{MAX}$  & $Q_{MAX}$      & $Q_{MIN}$      &  \\
					\hline
					32    & (9, 8, 10, 5)  & (9, 8, 10, 5) & (6, 2, 7, 2) &  \\
					33    & (9, 8, 10, 5)  & (9, 8, 10, 5) & (6, 2, 7, 2) &  \\
					34    & (9, 8, 10, 5)  & (9, 8, 10, 5) & (6, 2, 7, 2) &  \\
					35    & (9, 8, 10, 5)  & (9, 8, 10, 5) & (6, 2, 7, 2) &  \\
					36    & (9, 8, 10, 5)  & (9, 8, 10, 5) & (6, 2, 7, 2) &
				\end{tabular}}
			\end{minipage}
			\caption{The location on the lattice of global minima and maxima in the \emph{conp0050} configuration.  Changes between cooling iterations are highlighted in \textbf{bold}.}
			\label{tab::mv_case_study_2_field_maxima}
		\end{table*}
		
The stability of the lattice is also reflected by monitoring the location of minima and maxima in the topological charge density $Q$ and Wilson action $S$ fields (Table~\ref{tab::mv_case_study_2_field_maxima}).  In this interval of the cooling process the predicted locations of field minima and maxima are extremely stable, indicating that the same object is persisting throughout.  The location of maxima in the Wilson action $S_{MAX}$ and topological charge density $Q_{MAX}$ coincide predicting the presence of an \antiInstanton{}.  

\subsection{Input fields}

\noindent
In order to use the 4D topological charge density with the \jcn{} algorithm~\cite{meta2015} the dimensionality is reduced to 3D by slicing along the the $t$ axis.  Each \jcn{} in this case study is constructed from two neighbouring time-slices ($t_{n}$, $t_{n+1}$), additionally as the time axis is also periodic in \latticeQCD{} the \jcn{} ($t_{max(n)}$, $t_{min(n)}$) is a valid input configuration.  Hence, for the $16 \times 8$ ensemble, as used in this study, we preserve temporal periodicity by computing a \jcn{} for ($t_{8}$, $t_{1}$).

\subsubsection{Slab size parameter}

\begin{figure*}[!h]
	\centering
	
	
	\begin{minipage}{.06\textwidth}
		\vfill
		\caption*{\tiny{Topological charge density}}
		\includegraphics[width=\textwidth]{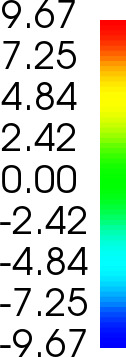}
		\vfill
	\end{minipage}
	\hfill
	\begin{minipage}{.3\textwidth}
		\includegraphics[width=\textwidth]{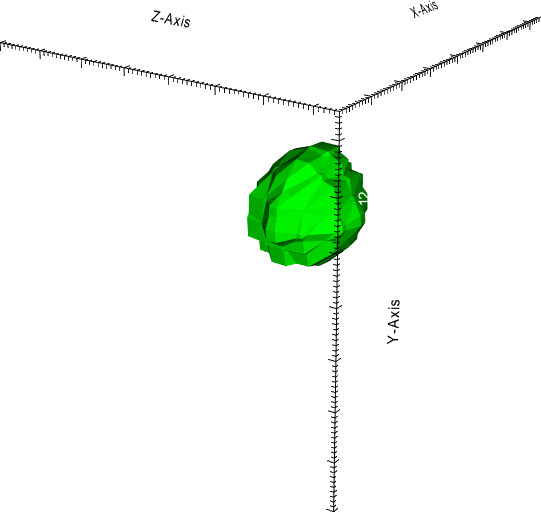}
	\end{minipage}
	\hfill
	\begin{minipage}{.3\textwidth}
		\includegraphics[width=\textwidth]{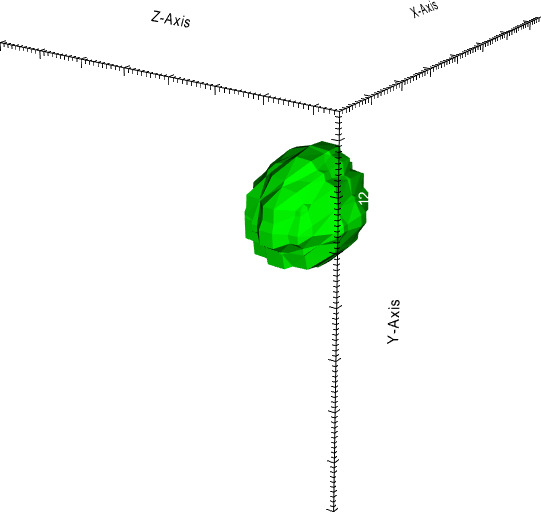}
	\end{minipage}
	\hfill
	\begin{minipage}{.3\textwidth}
		\includegraphics[width=\textwidth]{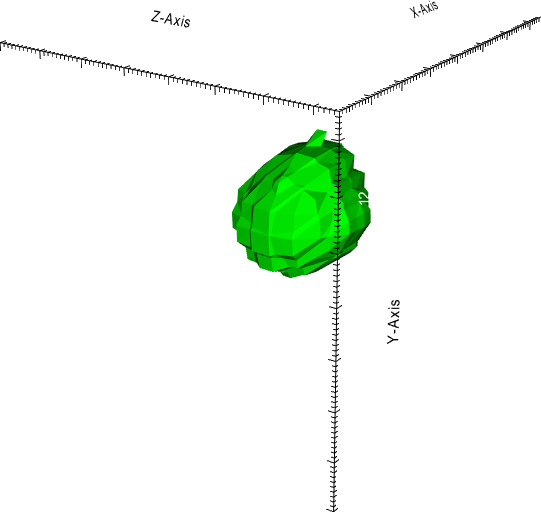}
	\end{minipage}
			
	\begin{minipage}{.06\textwidth}
	\end{minipage}
	\begin{minipage}{.3\textwidth}
		\caption*{$t = 4$}
	\end{minipage}
	\hfill
	\begin{minipage}{.3\textwidth}
		\caption*{$t = 5$\\(maxima)}
	\end{minipage}
	\hfill
	\begin{minipage}{.3\textwidth}
		\caption*{$t = 6$}
	\end{minipage}
	\caption{View of the main instanton observable as univariate slabs.}
	\label{fig::instanton_in_jcn}
\end{figure*}

\noindent
Throughout this study the \jcn{} slab size parameter is fixed dividing the topological charge density, with approximate range $[-45.0, 45.0]$, into $2^9 = 512$ intervals to give a slab size of $0.17578125$.  Figure~\ref{fig::instanton_in_jcn} visualises the instanton as 3 consecutive temporal objects, generated as interval contours at the specified slab size.  The green shades of each object, used to show neutral isovalues, show the outer layers of the instanton hide a dense core at the field maxima.  The same object is present in each time step; however, its structure constantly evolves with time.  Multivariate persistence measures can be used to capture this information at the specified slab size as demonstrated in Sec.~\ref{sec::multivariate_temporal_quantitative}.

\subsection{Global location of \antiInstanton{} observables}
\label{sec::global_observable_location}

\noindent
Figure~\ref{fig::temporal_global_view} presents the eight separate \jcn{}s created by evaluating each pair ($t_n, t_{n+1}$) of temporal fields.  A fixed colour transfer function is used, based upon on the peak magnitude in four-dimensions, in order to present a true representation of potential \antiInstanton{}s in the data.  This technique was chosen as \latticeQCD{} observables must be considered as global extrema in 4D, rather than as localised between two time steps.

\include{graphics/jcn_global_overview}

\begin{figure*}[!h]
	\begin{minipage}{.08\textwidth}
		\vfill
		\caption*{\tiny{Topological charge density}}
		\includegraphics[width=\textwidth]{colour_ramp_global}
		\vfill
	\end{minipage}
	\hfill
	\begin{minipage}{.85\textwidth}
		\begin{minipage}{\textwidth}
			\includegraphics[width=\textwidth]{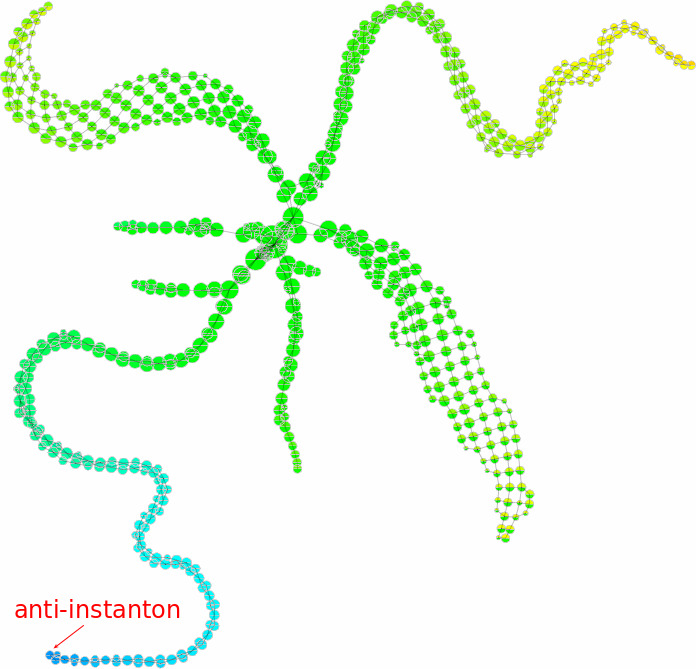}
		\end{minipage}
	\end{minipage}
	\caption{An anti-instanton can be located in the $t = (1, 2)$ \jcn{} as the branch with blue glyphs.  This object continues to exist for several time steps.}
	\label{fig::jcn_global_t1_t2}
\end{figure*}

Initially it is possible to locate an anti-instanton in the \jcn{} for $t = (1, 2)$ by identifying the object as a global minima using the coloured glyphs (Fig.~\ref{fig::jcn_global_t1_t2}).  The object can also be detected in $t = (2, 3)$ using the same approach; however, the bottom half of the glyph turns green to indicate a neutral isovalue.  The structure of the anti-instanton continues to persist in the data beyond this time slice despite the change in isovalue.  Recovery of the anti-instanton slab structure in later time slices requires some exploration of \jcn{} branches using knowledge of the objects location.  The anti-instanton becomes a prominent feature in the bivariate topology at $t = (8, 1)$ where the coloured glyphs indicate the presence of a global minima.

\begin{figure*}[!h]
	\begin{minipage}{.08\textwidth}
		\vfill
		\caption*{\tiny{Topological charge density}}
		\includegraphics[width=\textwidth]{colour_ramp_global}
		\vfill
	\end{minipage}
	\hfill
	\begin{minipage}{.85\textwidth}
		\begin{minipage}{\textwidth}
			\includegraphics[width=\textwidth]{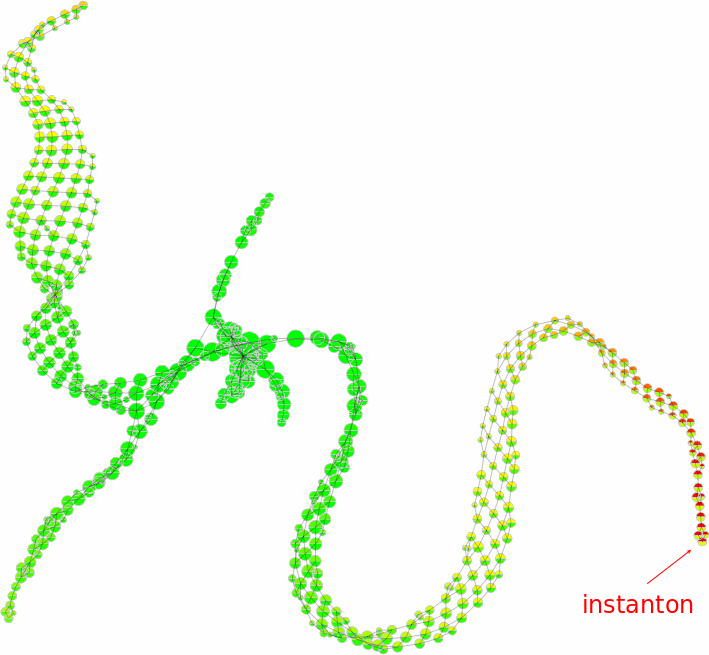}
		\end{minipage}
	\end{minipage}
	\caption{An instanton can be located in the $t = (5, 6)$ \jcn{} as the branch with red \& yellow glyphs.  The bottom half of the glyphs indicate that the topological charge density quickly drops off from a global peak at $t = 5$.}
	\label{fig::jcn_global_t5_t6}
\end{figure*}

Also present in the \jcn{} overview (Fig.~\ref{fig::temporal_global_view}) is a global maxima which first becomes prominent in the multivariate glyphs at $t = (4, 5)$.  This coincides with the output of the FORTRAN code which predicts the presence of an instanton at $t = 5$.  The instanton quickly begins to fade into the surrounding lattice structure at $t = (5, 6)$ where the multivariate glyphs indicate a return to more neutral isovalues (Fig.~\ref{fig::jcn_global_t5_t6}).

Several other localised features exist in the data at various points in the temporal \jcn{}s.  A second potential anti-instanton appears in the \jcn{} at $t = (7, 8)$, continuing to exist in parallel to the main anti-instanton observable at $t = (8, 1)$.  The \jcn{} offers an interesting approach to examining the interactions between multiple temporally localised lattice objects.  Lattice observables in neighbouring time slices could potentially have an influence over the geometric structure of other objects.


\subsection{Visually tracking an instanton across the temporal axis}
\label{sec::temporal_tracking}

\noindent
In the following section the bivariate topology, as captured by the \jcn{} and Reeb skeleton, is examined in greater detail.  We begin by looking at the identified global maxima in four dimensions, predicted by the cooling code as being at $(9, 8, 10, 5)$ and continue across the periodic boundary back to the origin.
%


%
%
%
%

\paragraph{Time steps $t = 4$ and $t = 5$}  The instanton is located by examining the \jcn{} vertices using the coloured glyphs relating to isovalue.  The \jcn{}, when drawn in domain layout, shows the approximate location of barycentre of the slabs making up the instanton.  Displaying the slab geometry makes it possible to validate that the location of the object agrees with the predicted location $(9, 8, 10, 5)$ from the cooling code.  The instanton structure is the most prominent region of the bivariate topology captured in the Reeb skeleton.  

%

Selecting Reeb skeleton vertices highlights common vertices in the \jcn{}.  Vertices in the Reeb skeleton share a one-to-many relationship with the \jcn{} as the Reeb skeleton collapses regions of path connected slabs into a single vertex.  In both fields the instanton is surrounded by slabs with low isovalue, represented by green glyphs \textemdash{} relating to the region of percolation on the lattice where \tcd{} centres on zero.  The Reeb skeleton facilitates the removal of the outer layers of the instanton by considering it as topological noise due to low persistence.  


%
%

\paragraph{Time steps $t = 5$ and $t = 6$}  The Reeb skeleton highlights two significant features for $t = (5, 6)$.  The most prominent feature is the instanton, but a secondary feature also stands out in the Reeb skeleton and \jcn{}.  Examination of the secondary object as slabs and using the \jcn{} shows a relatively small feature made up of a number of densely packed layers.  Isovalues of the object in both temporal fields reveal that the topological charge density at the inner core is neutral, indicated by the green coloured glyphs.  This allows it to be concluded that although a significant feature in the lattice topology this object is not a potential \antiInstanton{}.

\paragraph{Time steps $t = 6$ and $t = 7$}  The instanton is located using the \jcn{} as significant feature of the topology.  At $t = 6$ the instanton appears as a maxima in the topological charge density field; however, at $t = 7$ the isovalue of the instanton is drastically reduced to neutral.  The Reeb skeleton also identifies a potential anti-instanton, existing as a local minima at $t = 7$.  Also present is a secondary maxima that replaces the instanton structure as the local maxima at time slice $t = 7$.




\paragraph{Time steps $t = 7$ and $t = 8$}  The instanton structure continues to remain visible in the $t = (7, 8)$ \jcn{}.  The object becomes a less significant feature of the \jcn{} with green glyphs revealing that the isovalue has reduced to near zero.  However, enough topological structure remains to separate the instanton from the surrounding region of percolation.  Due to reduction of isovalue range representing the object the Reeb skeleton discards the instanton, instead determining it to be topological noise.

\begin{figure*}[!h]
	\centering
	\begin{minipage}{.9\textwidth}
		\includegraphics[width=.48\textwidth]{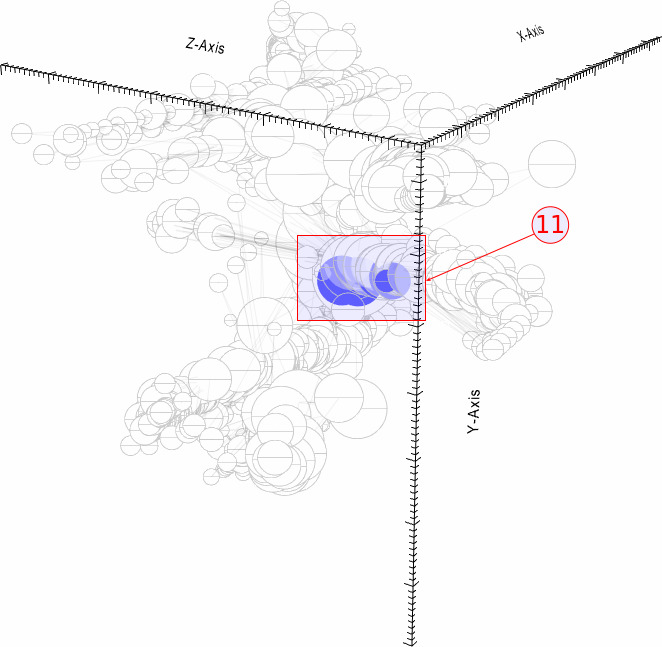}
		\hfill	
		\includegraphics[width=.48\textwidth]{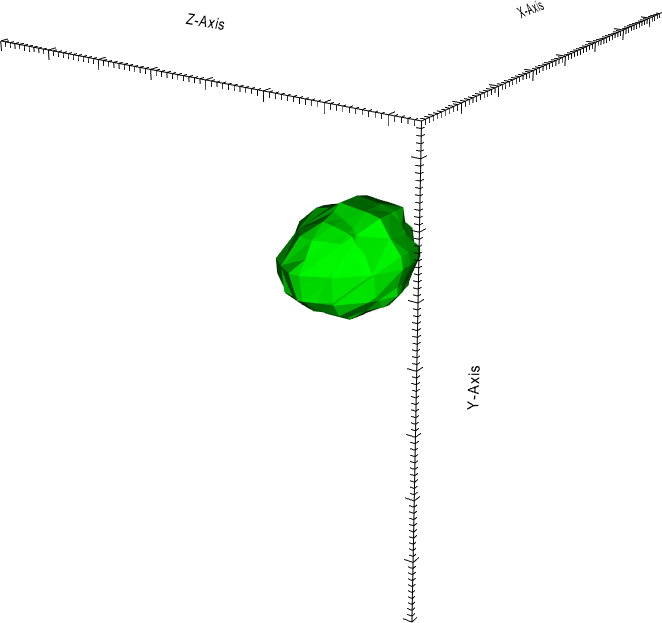}
	\end{minipage}
	\hfill
	\begin{minipage}{.08\textwidth}
		\vfill
		\caption*{\tiny{Topological charge density}}
		\includegraphics[width=\textwidth]{colour_ramp_global}
		\vfill
	\end{minipage}
	
	
	\caption{The \jcn{} in domain layout for $t = (8, 1)$ displayed alongside the joint contour slabs.  The selected glyphs (11) represent the instanton and correlate with the slab structure.}
	\label{fig::jcn_domain_t8_t1}
\end{figure*}

\paragraph{Time steps $t = 8$ and $t = 1$}  At standard resolution the \jcn{}, with inputs set to $t = 8$ and $t = 1$, failed to reveal any structure relating to the instanton observable.  However, when the slab size is decreased the instanton can be isolated from the surrounding topology charge density.  A halving of the slab size to $0.087890625$, giving $2^{10} = 1024$ intervals, is sufficient to allow the object to be located.  Examination of the geometric structure of the instanton, through the joint contour slabs, reveals that the shape found in earlier time slices begins to merge into the surrounding lattice field.  However, it is possible to confirm that the selected object relates to the instanton by observing the \jcn{} in domain layout (Figure~\ref{fig::jcn_domain_t8_t1}).

\paragraph{Time steps $t = 1$ and $t = 2$}  The \jcn{} for time-steps $t = (1, 2)$ reveals that the instanton can still be differentiated from the surrounding topological charge density by again halving the slab size to $0.087890625$.  Evidence of the emergence of a potential anti-instanton in the data can also be found in the \jcn{} and Reeb skeleton.  This coincides with the global minima $Q_{MIN}$ estimated by the cooling code to be present on the lattice at $(6, 2, 7, 2)$.


%

\paragraph{Time steps $t = 2$ and $t = 3$}  The main instanton observable reappears in the bivariate topology at standard resolution with inputs $t = (2, 3)$.  The \jcn{} and Reeb skeleton identify the instanton as a minor feature of the topology alongside several more significant features.  Two other features, interesting in the context of \latticeQCD{}, are present including the global minima in 4D \textendash{} a potential anti-instanton.  The presence of the anti-instanton appears not to distort the structure of the instanton.




\paragraph{Time steps $t = 3$ and $t = 4$}  The Reeb skeleton detects the instanton as the most prominent feature.  The slab structure captured by the \jcn{} is well defined and resembles that of other time slices.  The isovalues associated with the instanton object at $t = (3, 4)$ shows a large jump in isovalue between the two time-slices, captured by the colour change from red to green in the \jcn{}.  

\section{A quantitative approach to instanton tracking}
\label{sec::multivariate_temporal_quantitative}

\noindent
We have shown how the instanton can be tracked by the \jcn{} visually; however, for domain scientists a more quantitative approach is required.  In the following section we look at what statistical measurements are available using bivariate topological structures.  It should be noted that each of these process require manual locating of the target object.

\subsection{Slab count}

\noindent
A basic measure of persistence can be computed from the \jcn{} by counting the number of slabs making up a topological region.  Each vertex in the \jcn{} represents a joint contour slab, or region of the quantised Reeb space, associated with a pair of isovalues.  Throughout the case study it was found that the number of slabs linked to an object tended to vary with two properties of the object; the volume of the slabs, and the isovalue range of the object.  Sparse objects with a wide range of isovalues frequently appear as sheet-like structures in the \jcn{}, and densely packed objects appear with a branch-like structure.  

We found that it was possible to get a rough estimate of properties of lattice objects in the quantised Reeb space by calculating the percentage of vertices in the \jcn{} contributing to the object, as detected through manual selection.

\begin{table*}[!htb]
	\centering
	\begin{minipage}{0.48\textwidth}
		\resizebox{\textwidth}{!}{
		\begin{tabular}{llll}
		Input fields	& Instanton	 & Entire \jcn{} & Percentage \\
						& sub-graph	 & structure  & \\
		\hline
		$t = (1, 2)$ & $0$                   & $572$             & $0$                                \\
		$t = (2, 3)$ & $5$                   & $371$             & $1.35$                          \\
		$t = (3, 4)$ & $39$                  & $339$             & $11.50$                          \\
		$t = (4, 5)$ & $199$                 & $584$             & $34.08$                          \\
	\end{tabular}}
\end{minipage}
\hfill
\vline
\hfill
\begin{minipage}{0.48\textwidth}
	\resizebox{\textwidth}{!}{
		\begin{tabular}{llll}
		Input fields	& Instanton	 & Entire \jcn{} & Percentage \\
						& sub-graph	 & structure  & \\
		\hline
		$t = (5, 6)$ & $212$                 & $472$             & $44.92$                          \\
		$t = (6, 7)$ & $44$                  & $237$             & $18.57$                          \\
		$t = (7, 8)$ & $6$                   & $360$             & $1.67$                          \\
		$t = (8, 1)$ & $0$                   & $806$             & $0$                         \\
	\end{tabular}}
\end{minipage}
	\caption{Number of \jcn{} vertices contributing to instanton structure}
	\label{tab::slab_counts}
	\footnotesize{All values are taken using a slab size of $0.17578125$.}	
\end{table*}

\begin{figure*}[!htb]
	\centering
	\includegraphics[width=\textwidth]{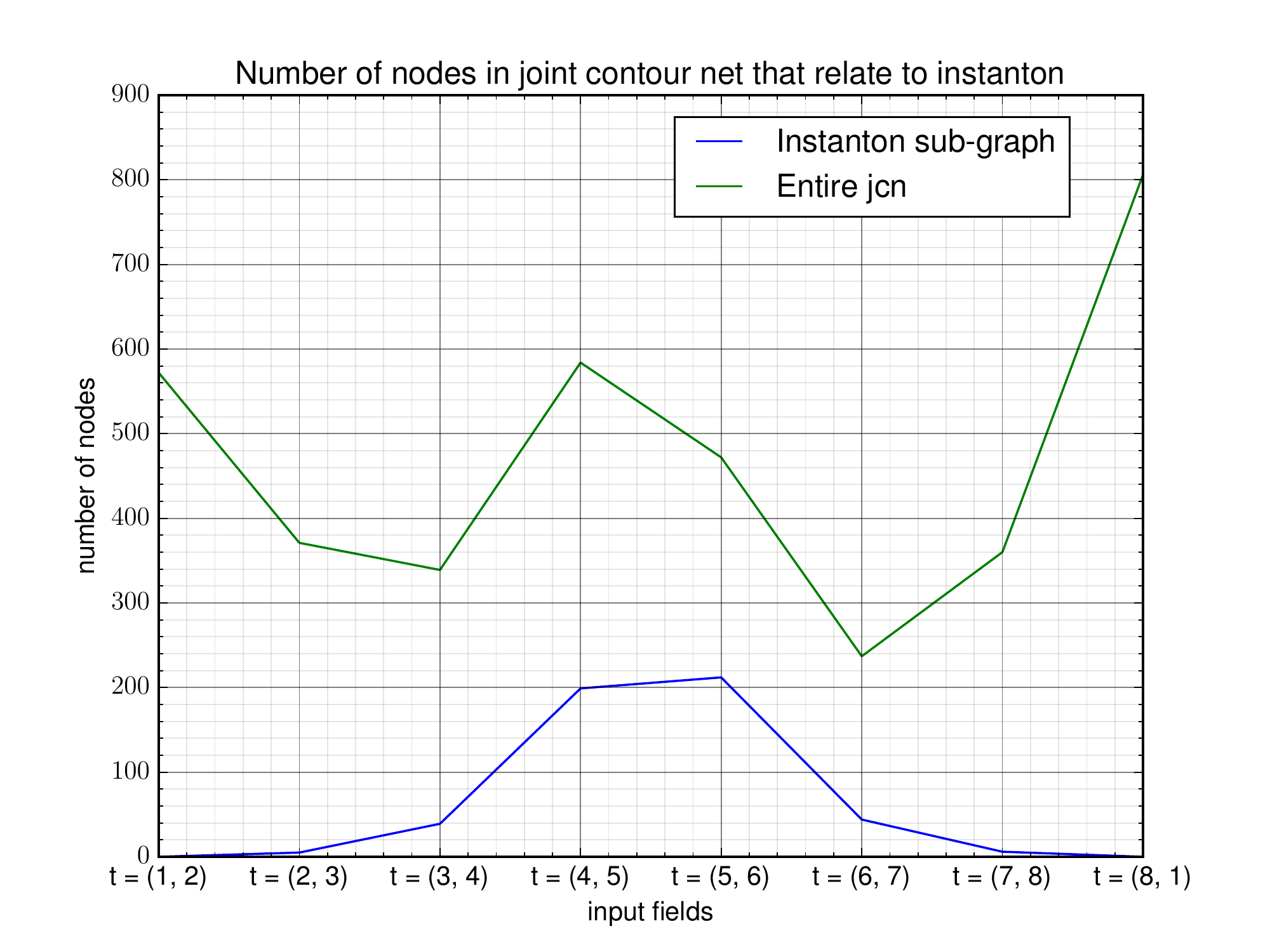}
	\caption{Number of slabs in the \jcn{} representing the instanton structure.}
	\label{fig::jcn_slabs_counts}
\end{figure*}

Table~\ref{tab::slab_counts} presents the number of vertices present in each \jcn{} across the temporal axis at the standard slab size.  When plotted as a histogram, as in Figure~\ref{fig::jcn_slabs_counts}, a peak in the number of vertices in the manually identified instanton sub-graph coincides with $Q_{MAX}$.  

The \jcn{}s show an increase in the number of total vertices on the lead up to $Q_{MAX}$ and $Q_{MIN}$, followed by a drop in graph complexity after each event.  The peaks were expected to coincide with the emergence of the anti-instanton and instanton; however, both peaks seem to proceed the time-steps containing the \antiInstanton{} structures rather than directly matching them.

\begin{figure*}[!htb]
	\centering
	\includegraphics[width=\textwidth]{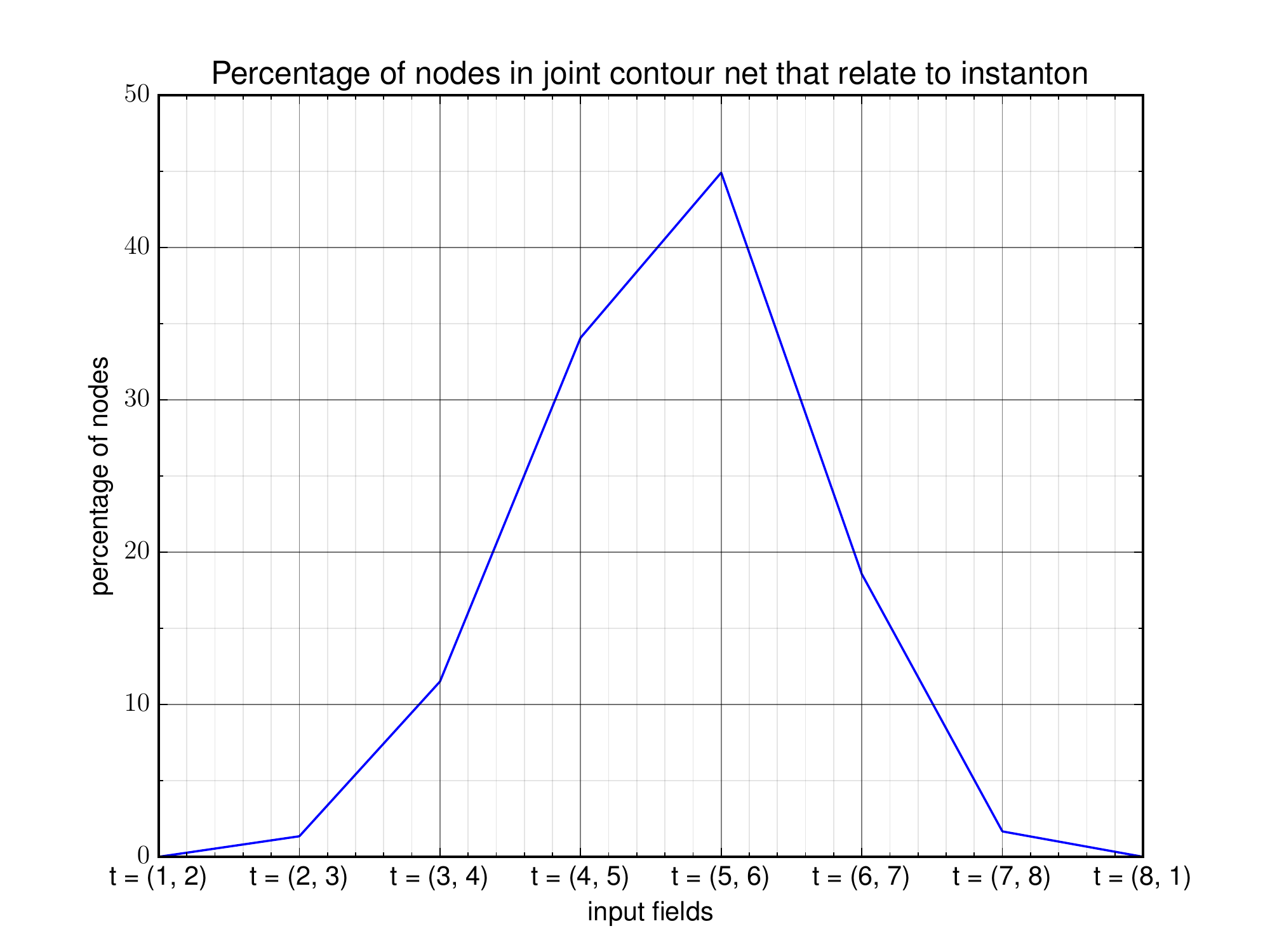}
	\caption{Percentage of slabs in the \jcn{} representing the instanton structure.}
	\label{fig::jcn_slabs_percentage}
\end{figure*}

Viewing the number of vertices in the instanton sub-graph as a percentage of the entire \jcn{} (Fig.~\ref{fig::jcn_slabs_percentage}) suggests that the structure dominates the quantised Reeb space at its peak value, where $t = 5$, agreeing with the predicted global maxima $Q_{MAX}$.  In the $t = (1, 2)$ and $t = (8, 1)$ \jcn{}s the percentage of vertices is zero \textemdash{} this is where the instanton structure was not detected at all at the 
standard resolution.


\subsection{Simplification and persistence data using the Reeb skeleton}
\label{sec::multivariate_simplification}

\noindent
The Reeb skeleton summarises the multivariate topology in a more compact graph structure.  Adjacent slabs are merged, provided no critical events coincide with the slab.  This allows entire path connected regions to be summarise as a single unit, represented by a vertex in the Reeb skeleton.  Besides simplification of the quantised Reeb space, the Reeb skeleton enables the generalisation of persistence measures used in univariate topology to multivariate topology.  Each vertex in the Reeb skeleton has persistence measures attached representing the connected components (the joint contour slabs), allowing us to further analyse the instanton structure.

This section examines how different persistence measures, used during simplification, affect the ability of the Reeb skeleton to detect the instanton structure.  The simplification measures are defined as follows:

\paragraph{Reeb skeleton}  Collapses adjacent slabs into a single vertex, meaning entire sheet or branch-like structures in the \jcn{} can be summarised by a single branch of the Reeb skeleton.  Topological events, such as splits and merges, are captured as vertices with degree 3 or higher.  No simplification is performed, frequently leading to the creation of multiple leaf vertices along branches of the Reeb skeleton.

\paragraph{Simplified Reeb skeleton}  This is the full Reeb skeleton except a basic pruning of non-critical vertices is performed.  First, all degree one singular vertices (leaf nodes) are removed from the Reeb skeleton and replaced with regular vertices.  Next any degree two singular vertices are replaced with regular nodes.  Finally, any regular leaf vertices are merged with their neighbours until a singular vertex is encountered.  This has the effect of collapsing large branches of regular vertices to a single vertex representing the entire set of connected components.

\paragraph{Persistence simplified Reeb skeleton}  The Reeb skeleton is first pruned of non-critical vertices, as described above.  Following this each of the remaining components in the graph are assigned a level of persistence based upon the quantity of \jcn{} nodes that make up the sub-graph represented by the vertex.  Next regular vertices of the \jcn{} that fall below a specified threshold are removed from the graph.  Finally, any remaining non-critical vertices are removed by repeating the basic pruning technique.

\paragraph{Volume simplified Reeb skeleton}  This method of simplification is similar to \quotes{persistence simplification} except instead of counting the number of \jcn{} nodes in a sub-graph of connected components, the volumes are approximated by counting the number of fragments passed through.  Each fragment represents a tetrahedra cell in the quantised Reeb space, counting the number of these in each slab gives a rough estimate of volume.

\begin{figure*}[!h]
	\centering
	\begin{minipage}{.75\textwidth}
		\includegraphics[width=\textwidth]{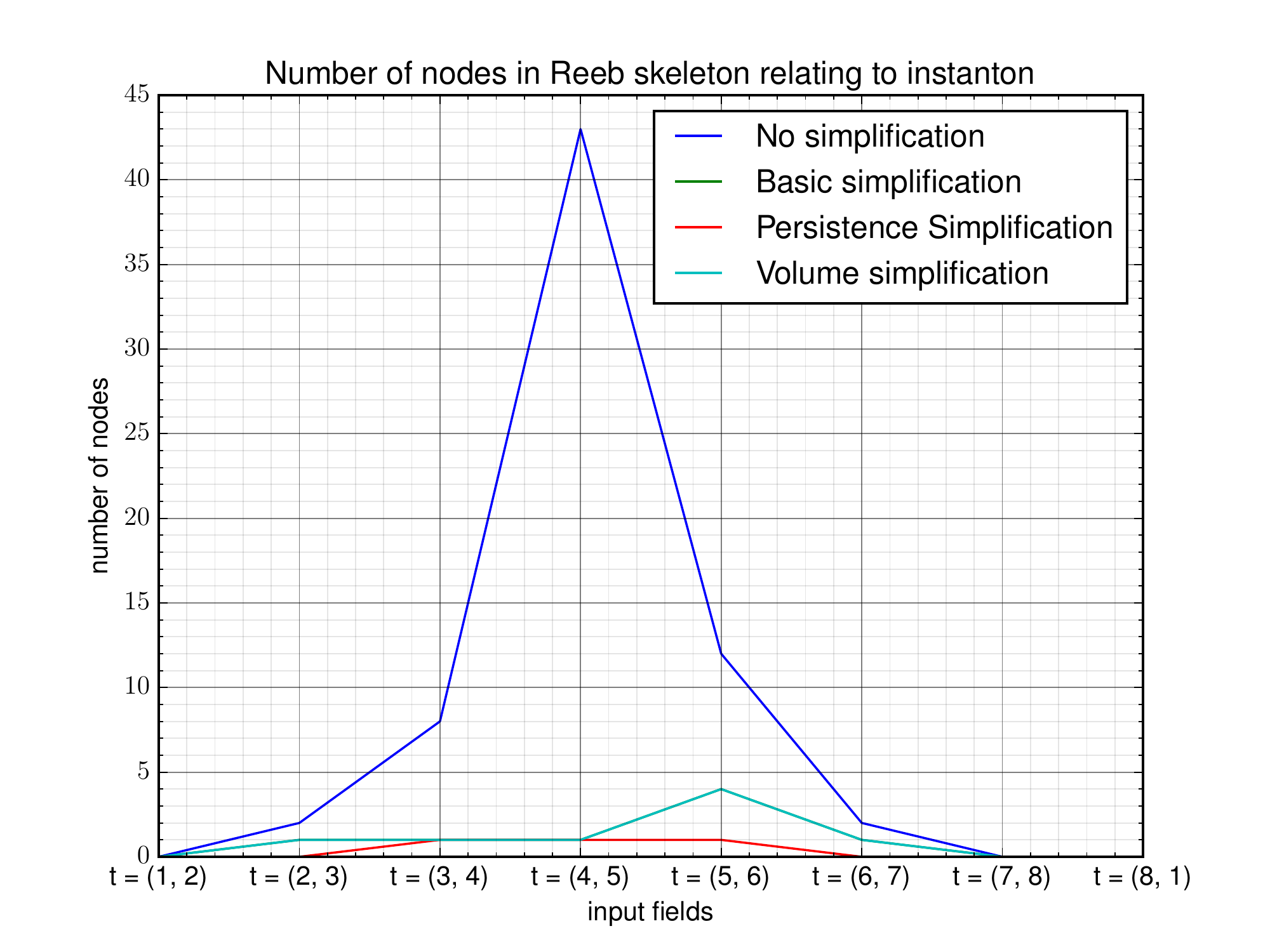}
		\caption{Number of nodes in the \emph{Reeb skeleton} representing the instanton under simplification.}
		\label{fig::reeb_skeleton_simplification}
	\end{minipage}
\end{figure*}

\noindent
Table~\ref{tab::reeb_skeleton_simplification} shows the effect that various forms of simplification have on the Reeb skeleton, also visualised in Figure~\ref{fig::reeb_skeleton_simplification}.  This confirms that under simplification the Reeb skeleton often reduces regions of the bivariate topology representing the instanton to a single vertex.

\begin{table}[!htb]
	\centering
	\resizebox{\textwidth}{!}{
	\begin{tabular}{lllllllll}
		& $t = (1, 2)$ & $t = (2, 3)$ & $t = (3, 4)$ & $t = (4, 5)$ & t = $(5, 6)$ & $t = (6, 7)$ & $t = (7, 8)$ & $t = (8, 1)$ \\
		\hline
		Unsimplified Reeb skeleton & $0$          & $2$          & $8$          & $43$         & $12$         & $2$          & $0$          & $0$          \\
		Simplified Reeb skeleton   & $0$          & $1$          & $1$          & $1$          & $4$          & $1$          & $0$          & $0$          \\
		Persistence simplification & $0$          & $0$          & $1$          & $1$          & $1$          & $0$          & $0$          & $0$          \\
		Volume simplification      & $0$          & $1$          & $1$          & $1$          & $4$          & $1$          & $0$          & $0$         
	\end{tabular}}
	\caption{Number of vertices in Reeb skeleton after simplification}
	\label{tab::reeb_skeleton_simplification}
\end{table}

\begin{table}[!htb]
	\centering
	\resizebox{\textwidth}{!}{
		\begin{tabular}{lllllllll}
			& $t = (1, 2)$ & $t = (2, 3)$ & $t = (3, 4)$ & $t = (4, 5)$ & $t = (5, 6)$ & $t = (6, 7)$ & $t = (7, 8)$ & $t = (8, 1)$ \\
			\hline
			Manual selection		   & $0$ & $5$ & $39$ & $199$ & $212$ & $44$ & $6$ & $0$ \\
			Unsimplified Reeb skeleton & $0$ & $5$ & $34$ & $186$ & $212$ & $28$ & $0$ & $0$ \\
			Simplified Reeb skeleton   & $0$ & $2$ & $33$ & $167$ & $122$ & $1$  & $0$ & $0$ \\
			Persistence simplification & $0$ & $0$ & $26$ & $137$ & $120$ & $0$  & $0$ & $0$ \\
			Volume simplification      & $0$ & $2$ & $33$ & $167$ & $122$ & $1$  & $0$ & $0$         
		\end{tabular}}
		\caption{Number of node in \jcn{} relating to instanton after simplification}
		\label{tab::jcn_under_simplification}
	\end{table} 

\begin{figure*}[!h]
	\centering
	\begin{minipage}{\textwidth}
		\includegraphics[width=\textwidth]{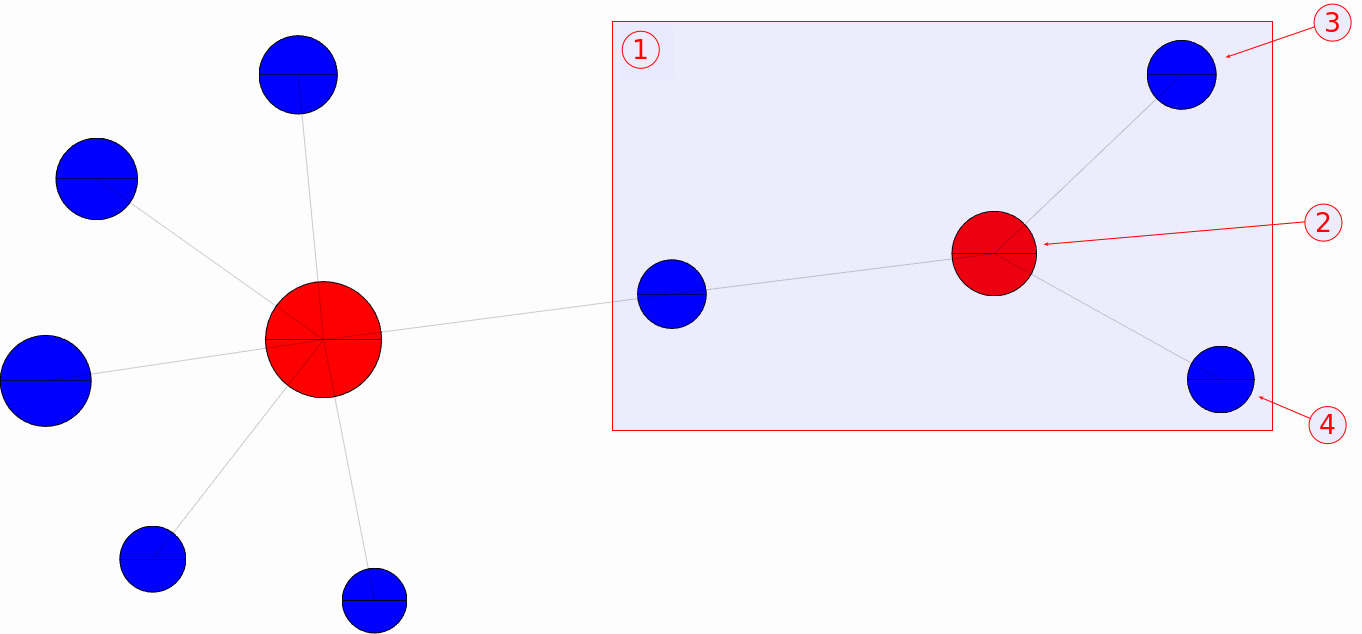}
		\caption{Reeb skeleton for $t = (5, 6)$, simplified using volume persistence measures.  The instanton (1) is captured as a branch in the simplified Reeb skeleton.  At vertex (2) the instanton splits into two; the outer part (3) and inner core (4).}
		\label{fig::forking_behaviour}
	\end{minipage}

	\begin{minipage}{\textwidth}
		\begin{minipage}{.08\textwidth}
			\vfill
			\caption*{\tiny{Topological charge density}}
			\includegraphics[width=\textwidth]{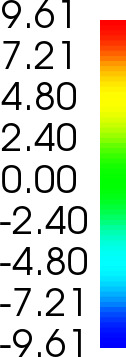}
			\vfill
		\end{minipage}
		\hfill
		\begin{minipage}{.9\textwidth}
			\includegraphics[width=.32\textwidth]{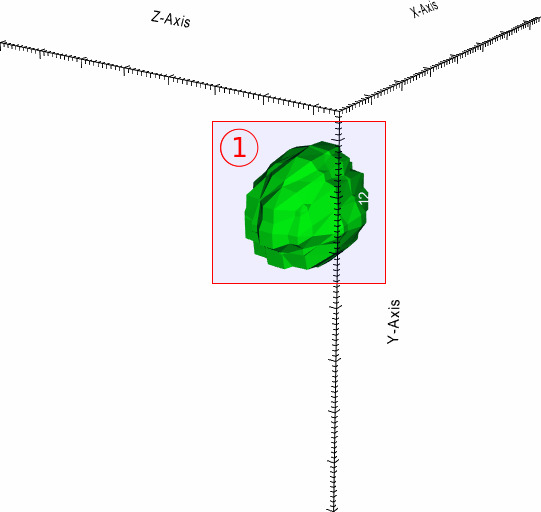}
			\hfill
			\includegraphics[width=.32\textwidth]{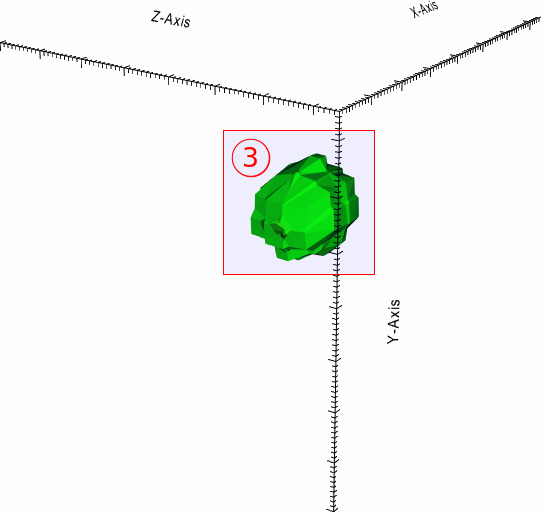}
			\hfill
			\includegraphics[width=.32\textwidth]{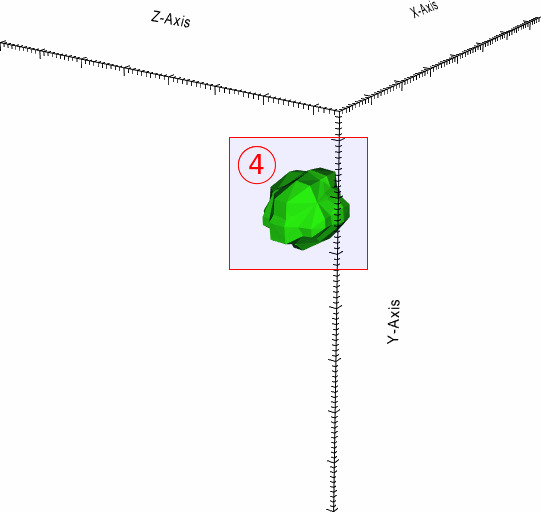}
		\end{minipage}		
	
		\begin{minipage}{.08\textwidth}	
		\end{minipage}
		\hfill
		\begin{minipage}{.3\textwidth}	
			\caption*{Slab structure}
		\end{minipage}
		\hfill
		\begin{minipage}{.3\textwidth}	
			\caption*{Outer core}
		\end{minipage}
		\hfill
		\begin{minipage}{.3\textwidth}	
			\caption*{Inner core}
		\end{minipage}
		
		\caption{The forking behaviour visualised using the slab geometry.  The full instanton structure (1) splits into outer (3) and inner (4) parts.}
		\label{fig::forking_slabs}
	\end{minipage}
		
\end{figure*}

\begin{figure*}[!htb]
	\centering
	\begin{minipage}{.75\textwidth}
		\includegraphics[width=\textwidth]{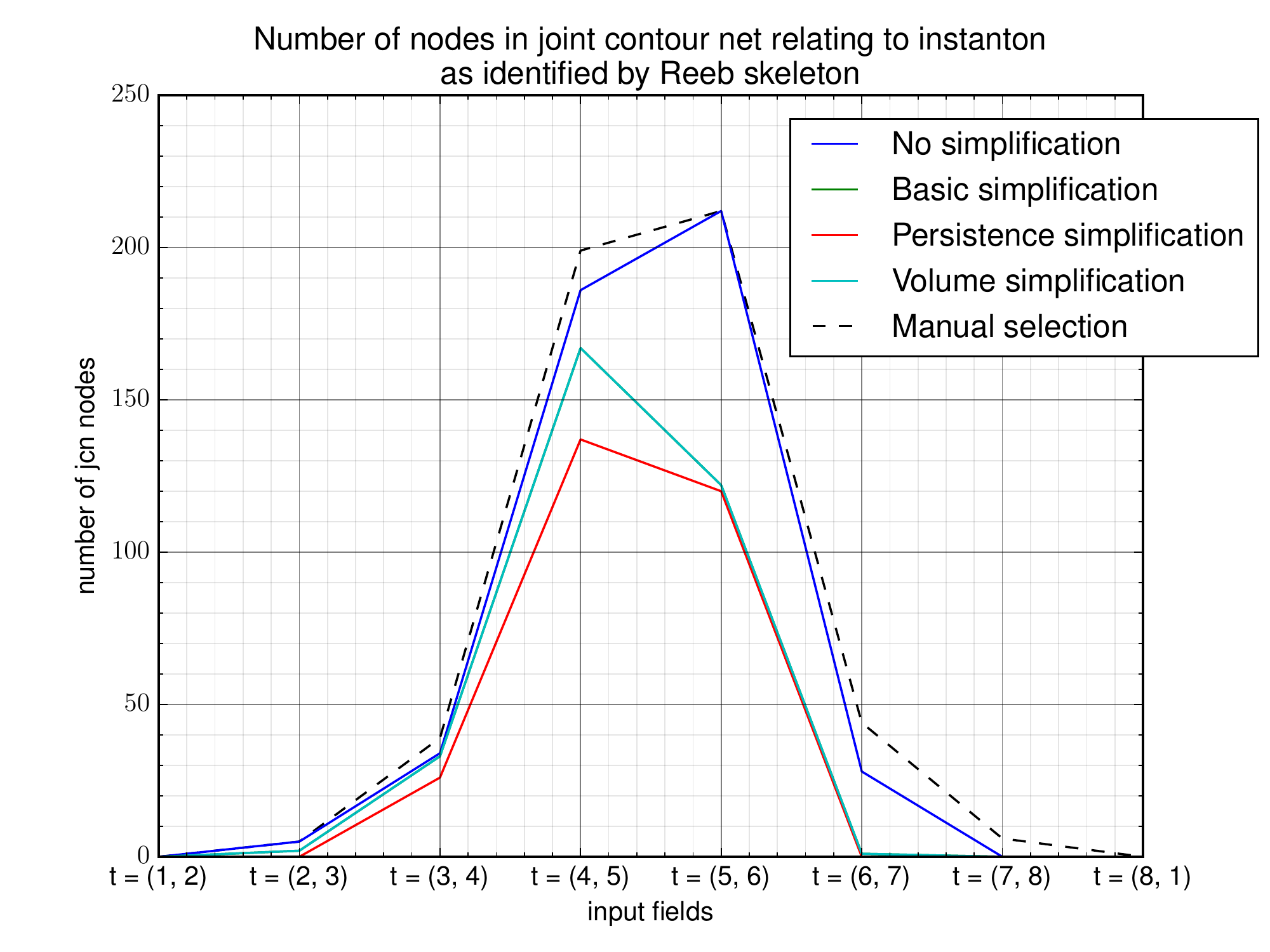}
		\caption{Number of nodes in the \emph{\jcn{}} representing the instanton under simplification.}
		\label{fig::jcn_under_simplification}
	\end{minipage}
\end{figure*}

Simplification using basic leaf pruning and volume measures at first appear to give similar results.  However, when viewing the effect by cross referencing the \jcn{} (Table~\ref{tab::jcn_under_simplification}, Fig.~\ref{fig::jcn_under_simplification}) variations appear, as smaller slabs are filtered out under simplification.  The simplification preserves the observed forking behaviour in the instanton structure at $t = (5, 6)$, resulting in four nodes in the Reeb skeleton (Fig.~\ref{fig::forking_behaviour}). Investigating the object by viewing the slab geometry (Fig.~\ref{fig::forking_slabs}) shows the instanton splits into an outer (3) and inner (4) shell structure.

Figure~\ref{fig::jcn_under_simplification} shows how under simplification the number of vertices in the \jcn{} judged to be topological noise varies by the persistence measure used.  The unfiltered Reeb skeleton captures most of the vertices that are chosen by manual selection.  Minor variations are found at the point where the instanton branch of the Reeb skeleton attaches to the region of percolation around zero.  This appears as a very highly connected vertex at the centre of the \jcn{} and Reeb skeleton.  Under simplification most techniques merge the \jcn{} vertices closest to isovalue zero into the central vertex due to their low relative persistence.  

\begin{figure*}[!htb]
	\centering
	\begin{minipage}{.32\textwidth}
		\includegraphics[width=\textwidth]{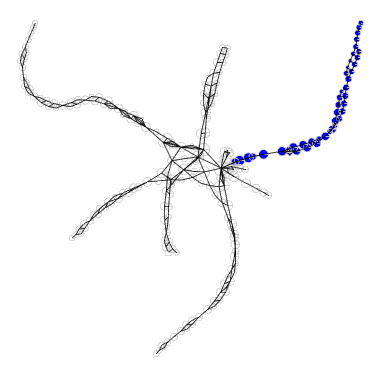}
	\end{minipage}
	\hfill
	\begin{minipage}{.32\textwidth}
		\includegraphics[width=\textwidth]{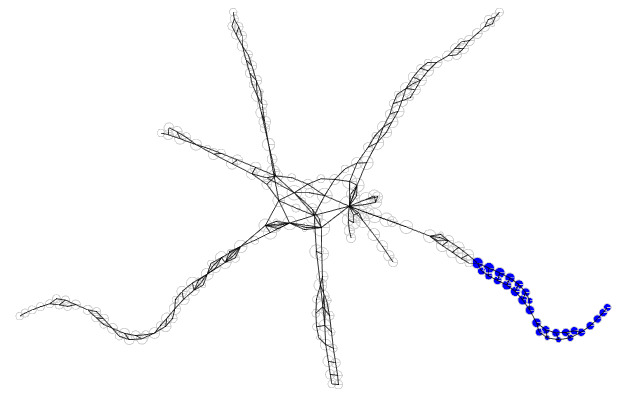}
	\end{minipage}
	\hfill
	\begin{minipage}{.32\textwidth}
		\includegraphics[width=\textwidth]{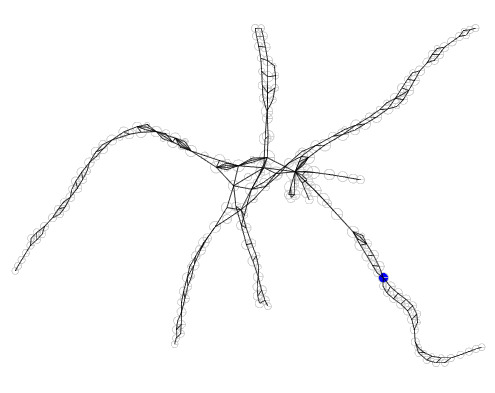}
	\end{minipage}
	
	\begin{minipage}{.32\textwidth}
		\caption*{Manual selection.}
	\end{minipage}
	\hfill
	\begin{minipage}{.32\textwidth}
		\caption*{Unsimplified Reeb skeleton selection.}
	\end{minipage}
	\hfill
	\begin{minipage}{.32\textwidth}
		\caption*{Volume simplified Reeb skeleton selection.}
	\end{minipage}
	\caption{Comparison of simplification techniques on the $t = (6, 7)$ joint contour net.}
	\label{fig::simplification_comparison}
\end{figure*}

We also observed that the vertices at the ends of \jcn{} branches, where the isovalue is furthest from zero, representing the core of an \antiInstanton{} are often removed from the intended observable.  There were other situations where simplification resulted in a single disconnected joint contour slab halfway along the instanton branch (Fig.~\ref{fig::simplification_comparison}).  This is often linked to the creation of many small and low persistence slabs \textemdash{} most likely this is an indication that the slab size parameter is too high.

The most aggressive form of simplification witnessed in this case study was based upon the number of connected components.  This removed the intended instanton observable from the simplified topology several time-steps earlier than other simplification techniques.  We believe this was likely caused by an overly strict threshold level and highlights the need for careful selection of simplification parameters.  

\section{Feature detection using the entire lattice data}
\label{sec::multivariate_temporal_full_lattice}

\noindent
The final use of the \jcn{} is an experimental attempt to detect the instanton using all eight time-slices as input.  Whilst the Reeb space is only well defined for situations where $m > n$, the \jcn{} algorithm can be used on data where this condition is not met.  In this situation we will be dealing with data defined on an $m = 3$ dimensional mesh with $n = 8$ function values.  The output of the algorithm is a subdivision of the input field with regard to all 8 functions; embedded amongst this we expect to find structure relating to the instanton.

\begin{figure*}[!htb]
	\centering
	\begin{minipage}{.08\textwidth}
		\vfill
		\caption*{\tiny{Topological charge density}}
		\includegraphics[width=\textwidth]{colour_ramp_global}
		\vfill
	\end{minipage}
	\hfill
	\begin{minipage}{.9\textwidth}
		\includegraphics[width=\textwidth]{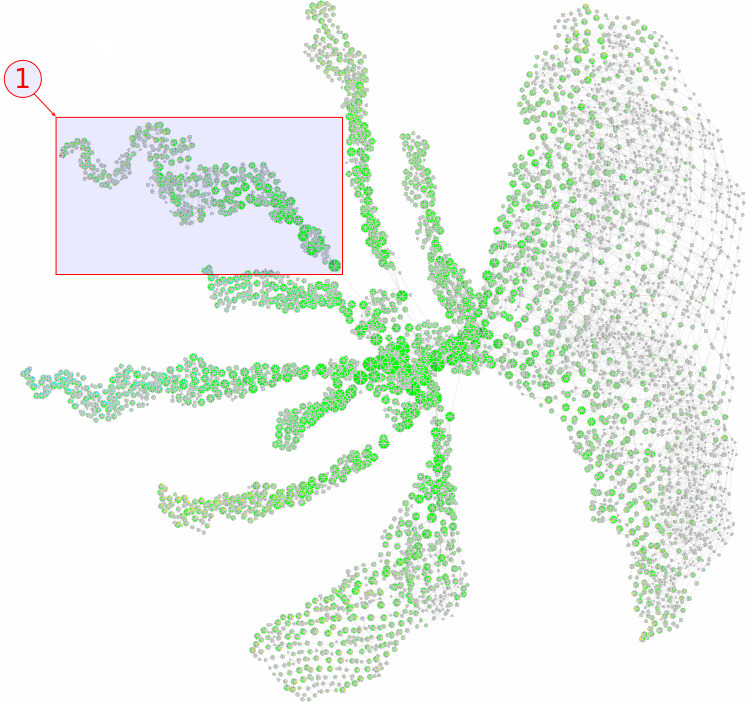}
	\end{minipage}

	\begin{minipage}{.8\textwidth}
		\includegraphics[width=\textwidth]{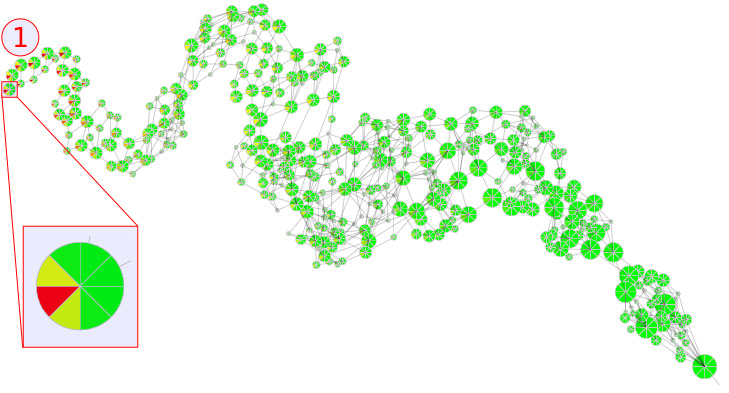}
	\end{minipage}
	\caption{Top: The \jcn{} for \mbox{$t = (1, 2, 3, 4, 5, 6, 7, 8)$.} captures the instanton as branch (1).  Bottom: Examining the branch in greater detail shows that the peak in topological charge density is capture in only a few vertices at the end of the branch.  The peak in the topological charge density is captured by the multivariate glyph at $t = 5$.  It is also possible to observe elevated levels in the topological charge density in neighbouring time slices.}
	\label{fig::t1t2t3t3t4t5t6t7t8_jcn}
\end{figure*}

\begin{figure*}[!htb]
	\centering
	\begin{minipage}{.08\textwidth}
		\vfill
		\caption*{\tiny{Topological charge density}}
		\includegraphics[width=\textwidth]{colour_ramp_global}
		\vfill
	\end{minipage}
	\hfill
	\begin{minipage}{.9\textwidth}
		\includegraphics[width=\textwidth]{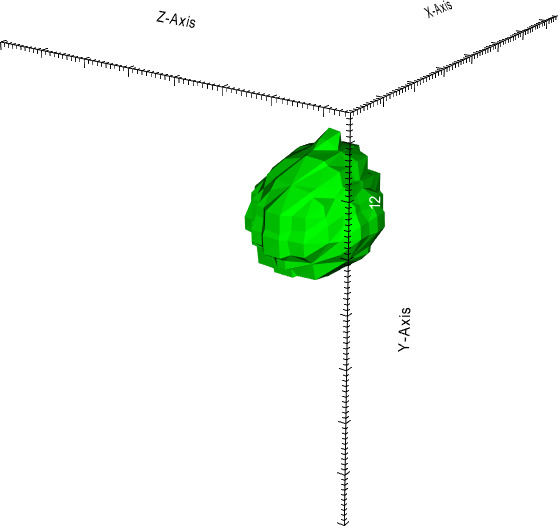}
	\end{minipage}
	\caption{The slab structure of the instanton recovered from the \jcn{} of \mbox{$t = (1, 2, 3, 4, 5, 6, 7, 8)$}.}
	\label{fig::t1t2t3t3t4t5t6t7t8_slabs}
\end{figure*}

The \jcn{} for $t = (1, 2, 3, 4, 5, 6, 7, 8)$ is shown in Figure~\ref{fig::t1t2t3t3t4t5t6t7t8_jcn}, featuring $36450$ vertices and $104373$ edges.  The structure clearly splits the input space into distinct topological objects, meaning it is possible to select a single branch of the \jcn{} to isolate the instanton structure (1).  Viewing the instanton as joint contour slabs (Fig.~\ref{fig::t1t2t3t3t4t5t6t7t8_slabs}) shows an object very similar to that output by the bivariate \jcn{}s.

Stepping through the data, colouring the graph by isovalue for each time-step, allows the variation in isovalues of each branch to be observed.  When reaching peak topological charge density $Q_{MAX}$ at time-step $t = 5$ all other vertices in the graph have very low relative isovalues, indicated by blue glyphs.  This is largely a side-effect of the colour transfer function being relative to the data set in view \textemdash{} for $Q_{MAX}$ this offsets the values further from zero than other time-steps.  The same effect is less obvious for $Q_{MIN}$ at $t = 2$, but can still be observed.  

\section{Conclusion}
\label{sec::multivariate_summary_temporal}

\noindent
We have shown in this paper how recent advances in multivariate topological analysis can be applied to \latticeQCD{} data sets.  We have demonstrated how multivariate topology can be used for comparing data with a temporal element in order to observe critical topological events as the time step is varied.  This case study also considered some basic multivariate persistence measures available in the data, such as the number of slabs.  These measures are suited to evaluation as lattice simulation parameters are varied as their values can be collected autonomously.

In summary this paper makes the following contributions:
\begin{itemize}
	\item The \jcn{} was used to track and observe the structure of an instanton in 4D space-time.
	
	\item Quantitative measures were taken directly from the \jcn{} and Reeb skeleton to evaluate the importance of lattice observables in the context of the topology of the lattice.
	
	\item We demonstrated how multivariate simplification metrics could potentially be utilised to locate important observables in \latticeQCD{}.
	
	\item It was also demonstrated how the \jcn{} can be used to reduce the structure of a 4D object to a 3D approximation.
\end{itemize}

\noindent
The study carried out in this paper forms part of a much larger body of work studying the use of topological visualisation techniques in lattice QCD~\cite{Solr-cronfa43497}.  The quantitative approaches demonstrated here can be used to perform analysis on ensembles of hundreds of configurations.  This is a typical use case for lattice QCD scientists, where averages across a large sample of configurations are needed to evaluate the effect of changing simulation parameters.  We give a more detailed review of how a domain expert may perform such a study in~\cite{Thomas2017}.

\section*{Acknowledgements}

\noindent
This work used the resources of the DiRAC Facility jointly funded by STFC, the Large Facilities Capital Fund of BIS and Swansea University, and the DEISA Consortium (www.deisa.eu), funded through the EU FP7 project RI-222919, for support within the DEISA Extreme Computing Initiative.  The work was also partly funded by EPSRC project: EP/M008959/1.


\bibliography{references}

\end{document}